\def\sbm{{\rm erg \, cm^{-2} \, s^{-1} \, arcmin^{-2} }}

\def\etal{{et al. }}

\def\keV{{\rm keV}}

\def\Omat{\Omega_{\rm M}}
\def\Olam{\Omega_{\Lambda}}

\documentstyle[emulateapj]{article}

\begin{document}
\lefthead{VOIT \& BRYAN}
\righthead{X-RAY SURFACE BRIGHTNESS DISTRIBUTION}
\slugcomment{\apj\  Letters, submitted 17 November 2000}
\title{On the Distribution of X-ray Surface Brightness from Diffuse Gas}
\author{G. Mark Voit\footnote{Space Telescope Science Institute,
3700 San Martin Drive, Baltimore, MD 21218, voit@stsci.edu} and
        Greg L. Bryan\footnote{Department of Physics, 
Massachusetts Institute of Technology, 
Cambridge, MA 02139}$^,$\footnote{Hubble Fellow}
}

\setcounter{footnote}{0}

\begin{abstract}
Hot intergalactic gas in clusters, groups, and filaments
emanates a continuous background of 0.5-2.0~keV X-rays that
ought to be detectable with the new generation of X-ray
observatories.  Here we present selected results
from a program to simulate the surface-brightness distribution
of this background with an adaptive-mesh cosmological hydrodynamics 
code.  We show that the bright end of this distribution is well
approximated by combining the cluster temperature function with
a $\beta$-model for surface brightness and appropriate 
luminosity-temperature and core radius-luminosity relations.
Our simulations verify that the X-ray background from hot gas
vastly exceeds observational limits if non-gravitational processes 
do not modify the intergalactic entropy distribution.  An
entropy floor $\sim 100 \, \keV \, {\rm cm}^2$, which could be 
established by either heating or cooling, appears necessary to
reconcile the simulated background with observations.
Because the X-ray background distribution is so sensitive to the
effects of non-gravitational processes, it offers a way to 
constrain the thermal history of the intergalactic medium 
that is independent of the uncertainties associated with 
surveys of clusters and groups.
\end{abstract}


\section{Introduction}

Beneath the scattered point sources that speckle the X-ray
sky should lie a subtler, more continuous background 
of X-rays emanating from the hot gas that fills the spaces 
between galaxies.  Many, if not most, of the universe's 
baryons inhabit intergalactic space and are heated to 
temperatures of $10^5 - 10^7$~K by gravitationally-driven 
shocks (e.g., Cen \& Ostriker 1999; Dav\'e \etal 2000).  
Once heated, these baryons 
tend to settle into gravitational potential wells and 
assume the characteristic temperature of the dark matter 
that confines them.  Because the emissivity of these
baryons depends on how severely they are compressed,
the mean intensity of the continuous X-ray background 
reflects the amount of non-gravitational energy 
injected into intergalactic space. Large amounts of energy
injection by supernovae and active galactic nuclei can
inhibit compression of the intergalactic medium, lowering
the mean level of the continuous background.

Current estimates place the point-source contribution to
the 0.5-2~keV background at $\gtrsim 80$\% (Hasinger \etal 
1998; Mushotzky \etal 2000; Giacconi \etal 2000),
meaning that hot intergalactic gas contributes less than
20\%, which amounts to $\lesssim 5 \times 10^{-16} \, \sbm$.  
A sizeable fraction of this residual emission comes from 
clusters of galaxies.  Integrating over the observed luminosity 
function of clusters, assuming no luminosity evolution, places 
the background from clusters hotter than 5~keV at $\sim 2.5 \times 
10^{-16} \, \sbm$.  However, these clusters cover only $\sim 1$\%
of the sky.  The remainder is covered by a confused patchwork
of groups and intercluster filaments whose properties depend
critically on non-gravitational processes.  If gravitational
processes alone were responsible for establishing the entropy
distribution of intergalactic gas, emission from groups 
and filaments would vastly overproduce the remainder of the 
0.5-2~keV background (Pen 1999; Wu, Fabian, \& Nulsen 2000).  
Somehow, the lowest-entropy, most compressible gas has been
eliminated.  Non-gravitational heating has been the most widely
studied way of establishing this entropy floor, but it is also
possible that low-entropy gas has been removed by radiative 
cooling and subsequent condensation (e.g., Bryan 2000).

Analyzing individual groups and filaments to determine the
impact of energy injection and cooling will not be easy.  Several factors
complicate the task of compiling unbiased samples of X-ray
emitting groups: Groups are low surface-brightness objects,
some apparent groups are chance superpositions of galaxies, 
and the potential wells of individual galaxies can strongly affect
a group's X-ray properties (see Mulchaey 2000 for a review).
Projection effects further complicate matters.  Virialized
objects with $kT > 0.5$~keV cover over a third of the sky,
making individual filaments very difficult to isolate and
creating a significant probability that one group will overlap
another somewhere along the same line of sight (Voit, Evrard,
\& Bryan 2000).  For example, about three higher-redshift 
$> 0.5$~keV groups are expected to lie within the projected 
virial radius of a 0.5~keV group at $z \approx 0.1$.

Because of these projection effects, statistical analyses
of the 0.5-2~keV surface brightness distribution should be
explored as an alternative way to characterize the lowest 
surface-brightness structures in the X-ray sky and perhaps
to gauge the impact of non-gravitational energy injection and cooling.  
We have begun to investigate this brightness distribution using both
hydrodynamical simulations and semi-analytical techniques
and report some of our early results in this letter.
Section~2 briefly describes our simulations of the X-ray surface
brightness distribution and shows how the high-brightness
end can be approximated with a semi-analytical model.
Section~3 demonstrates that non-gravitational processes
strongly influence this distribution function, and \S~4 
summarizes our results.
   
\section{Simulated Surface-Brightness Distributions}

Our analysis of the X-ray background owing to intergalactic
gas centers on the quantity $P(S)$, the probability that
a given line of sight through the universe will have a 
0.5-2~keV surface brightness $>S$.  The ideal computation
of $P(S)$ would involve a hydrodynamical simulation 
encompassing the entire observable universe, but
this is currently infeasible.  Instead, we have chosen to
simulate a box large enough to contain a fair sample of
clusters and groups, yet small enough to resolve the cores
of these virialized objects.  A companion paper (Bryan
\& Voit 2001) describes in more detail
these simulations, which employ an adaptive mesh refinement 
technique for the hydrodynamics (Bryan 1999; Bryan \& Norman 
1997; Norman \& Bryan 1999).  Here we will focus on 
simulations of a $50 \, h^{-1} \, {\rm Mpc}$ box with $128^3$
grid points and mesh refinement down to a minimum cell size
of $24 \, h^{-1} \, {\rm kpc}$.

\begin{figure}
\plotone{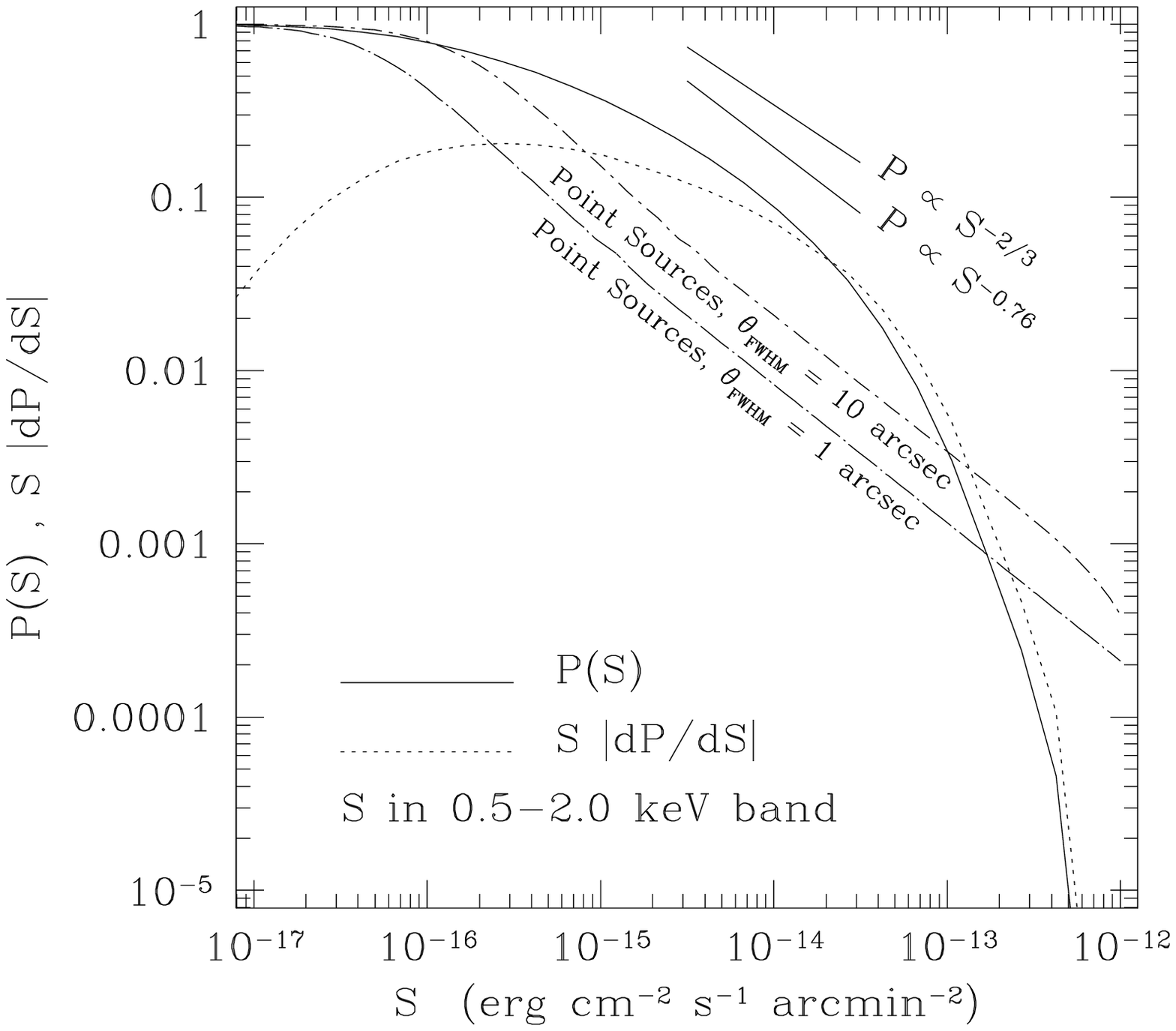}
\figcaption[sbplt_ps.ps]{Probability $P(S)$ of observing 
a 0.5-2.0 keV surface brightness $>S$ along a given line
of sight.  The solid line shows $P(S)$ for hot intergalactic
gas derived from a hydrodynamic simulation.  The dotted line
shows the corresponding differential distribution $S |dP/dS|$.
The other two lines show $P(S)$ owing to point sources for
two different Gaussian point-spread functions, one with
a full-width at half-max of 10 arcsec, the other with FWHM of
1 arcsec.
\label{sbplt_ps}}
\end{figure}

We reconstruct $P(S)$ for lines of sight from $z=0-10$ 
by computing $dP/dS$ for the box alone at a number of discrete
redshift points to determine how $dP/dS$ varies with redshift
for a given comoving box size.  At any given redshift, the 
comoving size of our simulation box corresponds to a redshift 
interval $\Delta z$, so we can reconstruct $dP/dS$ for the 
entire line of sight by appropriately convolving the individual 
distributions corresponding to each redshift interval (see Bryan
\& Voit 2001 for more details). This procedure yields
the distribution shown in Figure~\ref{sbplt_ps} for a $\Lambda$CDM
cosmology ($\Omat = 0.3$, $\Olam = 0.7$, $\sigma_8 = 0.9$) with
a baryon fraction $\Omega_b = 0.04$ and without non-gravitational 
energy injection.  Because of the limited box size, we cannot 
capture brightness enhancements owing to correlated structure
on $> 50 \, h^{-1} \, {\rm Mpc}$ scales, but we anticipate that
their effect on $P(S)$ will be small. 

Several features of this distribution are worth noting.  First,
the mean surface brightness is $\bar{S} = 2.5 \times 10^{-15} \, 
\sbm$, about five times higher than allowed by observations,
verifying the expectations of Pen (1999) and Wu \etal (2000).  
In addition, our numerical experiments reveal
that $\bar{S}$ depends on numerical resolution, implying
that $\bar{S}$ in an optimally resolved simulation would
be even higher (Bryan \& Voit 2001).  Second, $P(S) 
\approx 0.5$ at $S \sim 5 \times 10^{-16} \, \sbm$ indicating that
the median value of $S$ is close to the maximum allowed
by observations.  Third, the quantity $S |dP/dS|$ shows a
broad peak between $10^{-16}$ and $10^{-15} \, \sbm$, indicating
that most lines of sight have a surface brightness broadly
distributed in this range.  Finally, we show for comparison
two $P(S)$ distributions derived from the power-law fits to
ROSAT and Chandra point-source counts of Hasinger \etal (1998)
and Giacconi \etal (2000) assuming Gaussian point-spread
functions with full-width at half-max of 1 arcsec and
10 arcsec.  In both cases, diffuse hot gas dominates the
surface-brightness distribution from $10^{-16}$ to $10^{-13}
\, \sbm$.

An early hydrodynamical computation of X-ray surface brightness 
was performed by Scaramella, Cen, \& Ostriker (1993), who computed 
the mean and variance of the specific intensity at 1 and 2 keV for
a standard CDM cosmology using simulations of somewhat lower
effective resolution.  Because of the different cosmological
model and resolution limit, their results are difficult
to compare directly with our own.  However, we do verify their 
conclusions regarding the brightness distribution at intermediate
brightness levels.  Scaramella \etal (1993) found that the 
pixel distribution of specific intensity ($I_X$) varies like
$I_X^{-1.76}$, equivalent to $P(S) \propto S^{-0.76}$.  
Figure~\ref{sbplt_ps} illustrates this power-law slope, 
which is quite similar to that of our model in the neighborhood
of $10^{-14}$ to $10^{-15} \, \sbm$.

This scaling stems from the $\beta$-model surface brightness
distribution typical of clusters of galaxies.  Most clusters
are adequately fit by the law $S \propto [1 + (r/r_c)^2]^{-3\beta+1/2}$,
where $r_c$ is the cluster's core radius.
(Cavaliere \& Fusco-Femiano 1978).  For an individual cluster
we therefore have $P \propto S^{[2/(1-6\beta)]}$, which reduces to
$P \propto S^{-2/3}$ for $\beta = 2/3$, quite close to the
slope found by Scaramella \etal (1993). As Figure~\ref{sbplt_ps}
shows, this relation is also a good approximation of $P(S)$ 
in the intermediate range.

Because the simulations produce clusters and groups that are well
fit by $\beta$-models, we can successfully reproduce our simulated
$P(S)$ through semi-analytical means.  Following Voit \etal (2000),
we have taken the cluster catalog from the $\Lambda$CDM Hubble 
Volume simulation performed by the Virgo Consortium (Evrard 1999; 
Macfarland \etal 1998; Frenk \etal 2000) and have computed 
$P(S)$ assuming $\beta = 2/3$ surface-brightness profiles 
and the core radius-luminosity relation from Jones \etal (1998).  
However, instead of using the observed $L_X$-$T$ relation, we assume 
$L_{\rm bol} \propto T^2$, correct $L_X$ to the 0.5-2.0~keV band as described 
in Bryan \& Norman (1998) for a metallicity of 0.3 solar, and 
normalize the relation to fit our simulated clusters at $\sim 1$~keV.  
Figure~2 compares the resulting $P(S)$ distribution with the hydrodynamical 
model and the $P(S)$ derived from the Hubble Volume catalog using 
the {\em ROSAT} $L_X$-$T$ relation observed by Markevitch (1998)
instead of the one derived from the simulation.
   
Apparently, the contribution to $P(S)$ from the virialized
regions of clusters and groups can be adequately modeled by 
combining the cluster mass function with appropriate analytical
equations relating luminosity to mass and to a cluster's 
surface-brightness profile.  The excellent agreement between 
the simulated and semi-analytical $P(S)$ distributions also 
simplifies the task of identifying the major contributors at 
each level of $S$.  The steep slope of $P(S)$ at the 
surface-brightness levels of cluster cores ($\sim 10^{-13} 
\, \sbm $) echoes the steep slope of the cluster mass function.
Between $10^{-15} \lesssim S \lesssim 10^{-14} \, \sbm $ the 
$\beta$-model outskirts of clusters and groups dominate the background,
and below $10^{-15} \, \sbm $ is the realm of true intercluster
emission.

\section{The Signature of Non-Gravitational Processes}

\begin{figure}
\plotone{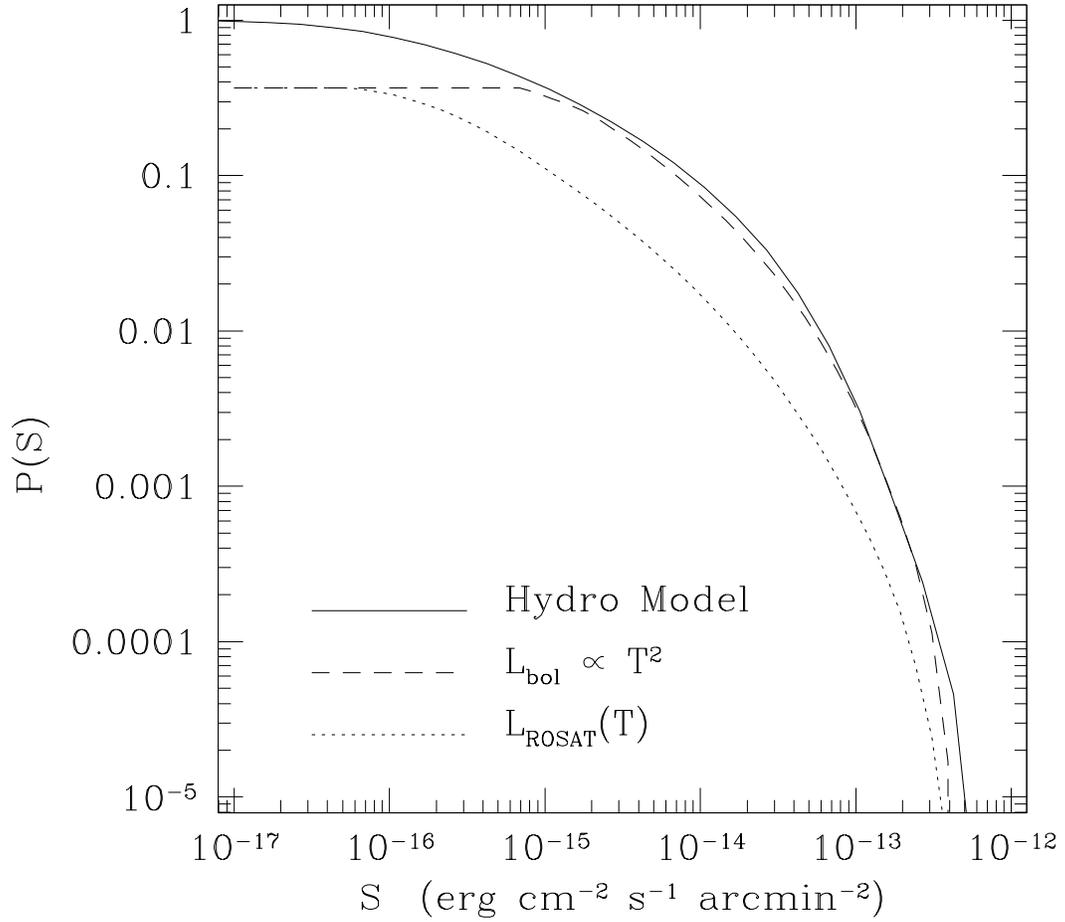}
\figcaption[sbplt_ps.ps]{Comparison of surface-brightness
distributions from hydrodynamical models and semi-analytical
methods.  Above brightness levels of $10^{-15} \, \sbm$ the $P(S)$
from our hydrodynamical model with no preheating (solid line)
is closely approximated by the $P(S)$ derived from the $\Lambda$CDM
Hubble-Volume simulation (dashed line) using a 
luminosity-temperature relation with $L_{\rm bol} \propto T^2$ and 
$\beta$-model surface-brightness profiles with the luminosity-core 
radius relation of Jones \etal (1998).  Appling the {\em ROSAT}
$L_X$-$T$ relation observed by Markevitch (1998) to the $\Lambda$CDM
Hubble-Volume simulation (dotted line) leads to a somewhat lower $P(S)$
distribution more consistent with current observational limits
on $\bar{S}$.
\label{sbcomp}}
\end{figure}

Our model without energy injection is illuminating but clearly
does not represent reality, primarily because $\bar{S}$ is far 
too high.  Applying the observed $L_X$-$T$ relation 
to the Hubble Volume clusters significantly shifts the $P(S)$ 
distribution to lower $S$ (see Figure~\ref{sbcomp}) with 
$\bar{S} \approx 6 \times 10^{-16} \, \sbm$, on the verge of 
being disallowed by observations.  However, this model relies
on an uncertain extrapolation of the cluster $L_X$-$T$ relation down 
to group scales.  In fact, the $L_X$-$T$ relation may steepen below 
$\sim 1 \, \keV$, perhaps because supernova energy injection 
becomes comparable to the gravitational energy of the
intragroup gas (e.g., Heldson \& Ponman 2000). 

\begin{figure}
\plotone{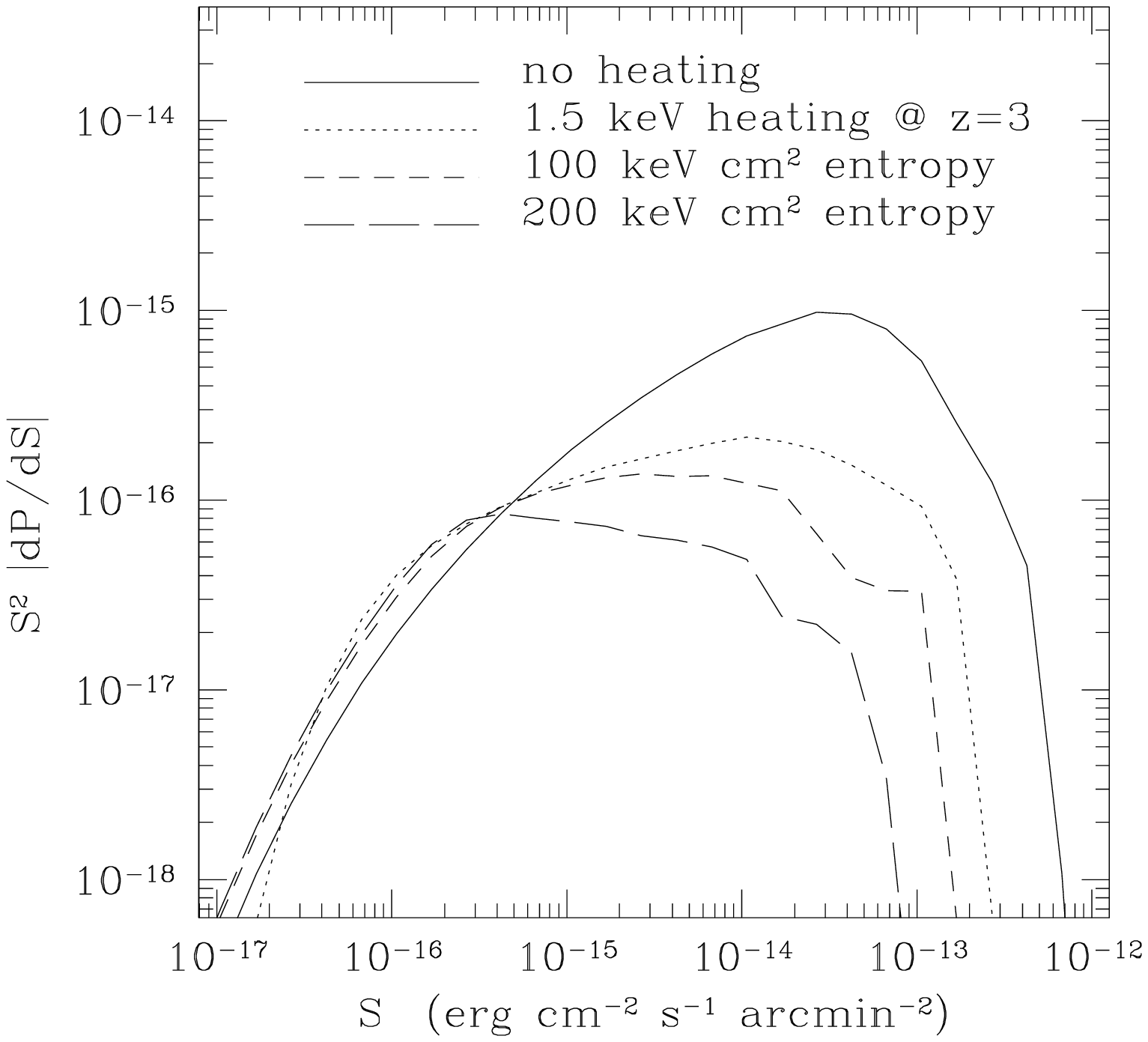}
\figcaption[sbplt_ps.ps]{Comparison of $S^2 |dP/dS|$ with various
treatments of non-gravitational processes.  The quantity $S^2 
|dP/dS|$ peaks at the surface-brightness levels that contribute 
most to the mean.  Without preheating or radiative cooling
(solid line), $S^2 |dP/dS|$ peaks at brightness levels typical 
of the cores of groups and clusters.  Adding energy equivalent 
to $1.5 \, \keV$ of energy per particle at $z = 3$ (dotted line) 
dramatically flattens $S^2 |dP/dS|$ and shifts it to lower $S$.
Applying an entropy floor to the simulation results (see text for 
details) has a similar effect.  Thus, observations of $S^2 |dP/dS|$
between $10^{-15}$ and $10^{-13} \, \sbm$ are potentially
very sensitive to the effects of non-gravitational processes.
\label{sbfbck}}
\end{figure}

In order to explore the effect of non-gravitational energy
injection on $P(S)$, we have run a hydrodynamical model with
a very simple prescription for preheating: we instantaneously
add $1.5 \, \keV$ of energy per baryon at $z=3$, similar to the
level needed to explain the observed $L_X$-$T$ relation for
clusters (e.g., Ponman, Cannon, \& Navarro 1999).  
Figure \ref{sbfbck} shows the resulting $P(S)$ 
in terms of the quantity $S^2 |dP/dS|$, which peaks in the 
neighborhood of $S$ values that contribute most to the mean.  
The distribution has indeed shifted to lower $S$, relative 
to the no-preheating case, and $S^2 |dP/dS|$ has also flattened, 
indicating that a larger range of $S$ contributes significantly 
to the mean.  Yet, the mean surface brightness, $\bar{S} = 
6.8 \times 10^{-16} \, \sbm $, still exceeds observational limits.
Furthermore, our numerical experiments indicate that even
this value may be an underestimate (Bryan \& Voit 2001).

Another crude but inexpensive way to way to explore the effects 
of non-gravitational processes is to apply an ad hoc entropy 
floor to the simulation results.  The heat input required
to explain the scaling properties of clusters corresponds to
a minimum entropy level $(Tn^{-2/3})_{\rm min} \sim 100 \, 
{\rm \keV \, cm^{2}}$ (e.g., Ponman \etal 1999).  
Interestingly, the same critical entropy level also emerges from 
cooling considerations:  $\sim 1 \, \keV$ gas will cool and
condense within a Hubble time if its specific entropy level
is $\lesssim 100 \, {\rm \keV \, cm^2}$.  In either
case, establishing an entropy floor lowers the mean gas density 
in the cores of groups and clusters.  Thus, we have chosen to
mimic the effect of an entropy floor by recalculating $P(S)$
after substituting the quantity $[n^{-2/3} + (Tn^{-2/3})_{\rm min}/T]^{-3/2}$
for the original gas density.  Figure~\ref{sbfbck} shows the
resulting $S^2 |dP/dS|$ distributions.  For $(Tn^{-2/3})_{\rm min}
= 100 \, {\rm \keV \, cm^2}$, we obtain $\bar{S} \approx 4.6
\times 10^{-16} \, \sbm$, just barely consistent with
current observations, and for $(Tn^{-2/3})_{\rm min}
= 200 \, {\rm \keV \, cm^2}$, we obtain $\bar{S} \approx 3.1
\times 10^{-16} \, \sbm$.

Croft \etal (2000) have recently performed a similar computation
using a smooth-particle hydrodynamics code that includes cooling and
a prescription for supernova feedback and find  
$\bar{S} = 6.4 \times 10^{-16} \, 
\sbm $.\footnote{After our letter was submitted, Phillips
\etal (2000) announced X-ray brightness results from an 
Eulerian hydrodynamical computation that includes cooling and
feedback in a $\Lambda$CDM cosmology with $\Omega_b = 0.035$.
Their value for the $0.5-2$~\keV background from diffuse gas,
$\bar{S} = 1.8 \times 10^{-16} \, \sbm$, is even lower, partly
because of the lower baryon fraction.}  Because the simulation
analyzed by Croft \etal includes multiple non-gravitational
processes that ours do not, it is difficult to pinpoint the process 
most responsible for lowering their value of $\bar{S}$.
Quite possibly, removal of low-entropy gas by cooling and 
condensation could be just as important as supernova heating 
in establishing $\bar{S}$ and the $L_X$-$T$ relation of groups 
and clusters (see also Bryan 2000).
It also remains possible that $\bar{S}$ would be larger in a 
higher-resolution SPH computation.  Additional simulations
will be needed to clarify the roles of various non-gravitational
processes and to relate the thermal history of intergalactic
baryons to the $P(S)$ distribution.

Given the difficulties in compiling an unbiased $L_X$-$T$
relation for groups, we suggest that observations of the $P(S)$
distribution at 0.5-2~keV be pursued as an alternative way
to constrain preheating and radiative cooling.  The quantity 
$S^2 |dP/dS|$ is clearly sensitive to non-gravitational processes 
like energy injection and cooling, particularly around $S \sim 
10^{-14} \, \sbm$, and the background from hot gas is competitive 
with the point-source background in {\em Chandra} and 
{\em XMM-Newton} observations over the interesting range of 
$S$ ($10^{-13} - 10^{-16} \, \sbm$).  Identification of extended 
sources and follow-up redshift surveys are unnecessary.  However,
the trick will be to distinguish the true astronomical background
from non-astronomical detector events.

\section{Summary}

We have derived the 0.5-2~keV surface-brightness distribution
function $P(S)$ from hydrodynamical simulations of structure
formation in a $\Lambda$CDM cosmology.  The bright end of this
distribution is well reproduced by combining the cluster
catalog from the $\Lambda$CDM Hubble Volume simulation with
appropriate $L_X$-$T$ and $r_c$-$L_X$ relations and a $\beta 
=2/3$ surface-brightness law.  Without non-gravitational
heating or cooling, $\bar{S}$ exceeds observed limits by
a factor of several.  Another simulation that adds $\sim 1
\, \keV $ of energy at $z=3$ substantially lowers $\bar{S}$,
but not by enough.  An entropy floor $\sim 100 \, \keV \, cm^2$
appears necessary to reconcile our simulations with observational
limits.  Observations of $P(S)$ with {\em Chandra}
and {\em XMM-Newton} appear to be a promising way to constrain
the thermodynamical history of this intergalactic gas.

\acknowledgements

GLB is supported by NASA through Hubble Fellowship
grant HF-01104.01-98A from the Space Telescope Science Institute,
which is operated under NASA contract NAS6-26555.  This research
has used data products made available by the Virgo
Consortium.


\newpage

\begin{references}

\reference{b99} Bryan, G. L. 1999, Computing in Science \&
   Engineering, March--April 1999, 46

\reference{b00} Bryan, G. L. 2000, \apj, in press

\reference{bn97} Bryan, G. L., \& Norman, M. L. 1997, in ASP Conf.
   Ser. 123, Computational Astrophysics, eds. D. A. Clarke \&
   M. J. West, (San Francisco: ASP), 1998, p. 363

\reference{bn98} Bryan, G. L., \& Norman, M. L. 1998, \apj,
   495, 80
   
\reference{cff78} Cavaliere, A., \& Fusco-Femiano, R. 1978,
   \astap, 70, 677
   
\reference{co99} Cen, R., \& Ostriker, J. P. 1999, \apj,
   514, 1

\reference{c00} Croft, R. \etal 2000, astro-ph/0010345

\reference{d00} Dav\'e, R. \etal 2000, astro-ph/0007217

\reference{e99} Evrard, A. E. 1999, in Evolution of large 
   scale structure: from recombination to Garching, 
   eds. A. J. Banday, R. K. Sheth, L.N. da Costa
   (Garching: ESO), p.249

\reference{hvsim} Frenk, C. S. \etal 2000, astro-ph/0007362

\reference{g00} Giacconi R. \etal 2000, astro-ph/0007420

\reference{h98} Hasinger, G. \etal 1998, \astap, 329, 482

\reference{hp00} Helsdon, S. F., \& Ponman, T. J. 2000, \mnras, 315, 356

\reference{j98} Jones, L. R., Scharf, C., Ebeling, H., Perlman, 
   E., Wegner, G., Malkan, M., \& Horner, D. 1998, \apj, 495, 100

\reference{mcpp99} MacFarland, T., Couchman, H. M. P., Pearce, F. R.,
   Pilchmeier, J. 1999, New Astronomy, 3, 687

\reference{m98} Markevitch, M. 1998, \apj, 504, 27

\reference{m00} Mulchaey, J. S. 2000, \araa, 38, 289 

\reference{mcba00} Mushotzky, R. F., Cowie, L. L., Barger, A. J., 
   Arnaud, K. A. 2000, Nature, 404, 459

\reference{nb99} Norman, M. L., \& Bryan, G. L. 1999, in Numerical
   Astrophysics: Proceedings of the International Conference on 
   Numerical Astrophysics 1998, eds. S. M. Miyama, K. Tomisaka, \&
   T. Hanawa (Boston: Kluwer), p. 19
   
\reference{p99} Pen, U. 1999, \apj, 510, L1

\reference{poc01} Phillips, L. A., Ostriker, J. P., \& Cen, R. 2000,
   astro-ph/0011348

\reference{pcn99} Ponman, T. J., Cannon, D. B., \& Navarro, J. F.
   1999, Nature, 397, 135

\reference{sco93} Scaramella, R., Cen, R. \& Ostriker, J. P. 1993,
   \apj, 416, 399

\reference{veb00} Voit, G. M., Evrard, A. E., \& Bryan, G. L. 2000,
   \apj, in press (astro-ph/0012191)
   
\reference{wfn00} Wu, K. K. S., Fabian, A. C., \& Nulsen, P. E. J.
  2000, \mnras, 318, 889
\end{references}
\end{document}